\documentclass[11pt,twoside]{article}
\usepackage{./asp2014}
\usepackage[utf8]{inputenc}
\usepackage{url}
\usepackage{amstext}

\aspSuppressVolSlug
\resetcounters

\begin{document}

\title{Modelling observable properties of rapidly rotating stars}
\author{Diego Casta\~{n}eda,$^1$ Robert G. Deupree,$^1$ and Jason Aufdenberg$^2$
\affil{$^1$Saint Mary's University, Halifax, Nova Scotia, Canada; \email{castaned@ap.smu.ca}}
\affil{$^2$Embry-Riddle Aeronautical University, Daytona Beach, Florida, USA; \email{aufded93@erau.edu}}
}

\paperauthor{Diego Castaneda}{castaned@ap.smu.ca}{0000-0001-8366-6905}{Saint Mary's University}{Astronomy and Physics}{Halifax}{Nova Scotia}{B3H 3C3}{Canada}
\paperauthor{Robert G. Deupree}{bdeupree@ap.smu.ca}{}{Saint Mary's University}{Astronomy and Physics}{Halifax}{Nova Scotia}{B3H 3C3}{Canada}
\paperauthor{Jason Aufdengerg}{aufded93@erau.edu}{}{Embry-Riddle Aeronautical University}{Physics}{Daytona Beach}{Florida}{32114}{USA}

\begin{abstract}
To fully understand the Be star phenomenon, one must have
a reasonable degree of knowledge about the star beneath the disk, which is often found to be rapidly rotating.
Rapid rotation complicates modelling because fundamental
properties like the stellar luminosity and effective temperature require knowledge of the
angle of inclination at which the star is observed. Furthermore our knowledge of the structure of rapidly rotating stars is
on a less sure foundation than for non-rotating stars. The uncertainties in the inclination and the surface properties
of a few rapidly rotating stars have been substantially reduced by interferometric observations over the last decade, and these stars 
can be used as tests of rotating stellar models, even if those stars themselves may not be Be stars.

Vega, as an MK standard, is historically a very important star because it is used for calibration purposes.
However, several studies have suggested that Vega is a rapidly rotating star viewed at a very low inclination angle, raising questions
as to how well we really know its properties. Appropriate modelling
has been challenging and there is still room for debate over the actual properties of Vega, as opposed to its observed properties.
We have previously shown that under certain conditions both the stellar surface properties and the
deduced surface properties scale from one model to another with the same surface shape.
We used this scaling algorithm with realistic 2D models to compute high-resolution spectral energy
distributions and interferometric visibilities to determine the best rotating model fit to Vega. Detailed comparisons between
the computed and observed data will be presented.
\end{abstract}

\section{Introduction}
Be stars tend to be the most rapidly rotating near main sequence stars with an average $V\,\mathrm{sin}\,i$ as much as 150 km s$^{-1}$
faster than similar B stars \citep{Slettebak1949}. Be stars are also more complex systems composed of a rapidly rotating star
surrounded by a disk, which makes modelling them much more
complicated. Cases like Achernar \citep{deSouza2003,Carciofi2008} have shown
that even distinguishing the stellar surface from the disk is challenging.

Realistic modelling of rotating stars may be an important factor in
understanding the Be phenomenon.
Advances in the general study of rotating stars, however, have been
limited by both theoretical and observational difficulties. A key
example is that fundamental properties such as the effective
temperature ($T_{\mathrm{eff}}$) and luminosity ($L$) that one would deduce
from observations now depend significantly on the angle of
inclination ($i$) between the line of sight and the star's rotation
axis for sufficiently rapidly rotating stars (e.g., \citealt{Collins1966,Hardorp1968,Maeder1970,Lovekin2006a,Gillich2008,Dall2011}).
This greatly complicates the determination of the star's position on the H-R diagram and hence
nearly all other useful information unless its inclination can be determined.

As for any kind of star, the usual information available for a rotating star consists of a
spectral energy distribution (SED),
broadband photometry and a deduced luminosity (deduced $L$).
The complication for rotating stars is that the SED is strongly dependent on the inclination.
Traditionally, analysis of the line profile broadening provides
a way to measure the projected rotational velocity ($V\,\mathrm{sin}\,i$) of the rotating star, but 
determination of the inclination and surface shape, however, are not easily measurable with standard observing methods.
Fortunately, important
advances in interferometric instrumentation over approximately
the last decade have permitted resolved observations of some
nearby rapid rotators (e.g., \citealt{vanBelle2001,deSouza2003,Aufdenberg2006,Monnier2007,Zhao2009,Che2011}).
This type of observations is able to measure $i$ as well as the surface shape of the star.

The strong dependence of the observed properties of a rotating star and its inclination requires a distinction between its
measured properties (i.e. deduced $T_{\mathrm{eff}}$ and deduced $L$) and its intrinsic properties, like the actual $L$,
 which is defined as the total energy leaving the star per unit time, and the actual $T_{\mathrm{eff}}$ which we define as
$(L/\left(A\sigma\right))^{1/4}$ where $A$ is the surface area of the star and $\sigma$ is the Stefan-Boltzmann constant.
The deduced $T_{\mathrm{eff}}$ is obtained applying non-rotating color - $T_{\mathrm{eff}}$ relations
to the observed colors of the star.

The direct process for finding appropriate models that fit the observations
requires performing the calculations with different models until
the observed SED properties are matched to the extent possible.
This can be laborious and it would be far preferable to be able
to start with the observed SED properties and work backward to
what the luminosity and the latitudinal variation of the effective
temperature must be. It is of great interest to find ways that can provide a bridge between
the deduced properties of a rotating star and its intrinsic properties.
In the next section we will discuss a methodology that may help with
this problem.

\section{Scaling of observed properties in rotating stars}
\label{sec:scaling}

\citet{Deupree2011a} showed that a number of properties
of rotating models, particularly the surface effective temperature
as a function of latitude, are proportional between models as long
as the surface shape 
(which we shall indicate by $R_{\mathrm{p}}/R_{\mathrm{eq}}$, although we mean 
having the same surface radius ratio applies at all latitudes) 
remains the same.
Furthermore, for the surface shapes to be exactly the same, the two models must
have the same rotation law to within a multiplicative constant.
The independence of latitude for the actual effective temperature
and radius ratios suggested that observable properties such as
the deduced luminosity and deduced effective temperature as
functions of inclination could scale as well. This scaling might
be able to at least place constraints on models and parameters
that could produce the observed properties. It would also allow
the deduction of the actual
luminosity and effective temperature from observations in a straightforward way
for cases in which both the inclination and surface shape are
known.

Recently, we showed \citep{Castaneda2014} that observable
properties of rotating stars that have the same shape are proportional
within each other. This effectively means that
\begin{equation}
\frac{\textrm{deduced }T_{\mathrm{{eff},}2}(\theta_{j})}{\textrm{deduced }T_{\mathrm{{eff},}1}(\theta_{j})}=c_{T_{\mathrm{eff}}}\quad\textrm{and}\quad\frac{\textrm{deduced }L_{2}(\theta_{j})}{\textrm{deduced }L_{1}(\theta_{j})}=c_{L}\quad\forall\quad j\label{eq:1}
\end{equation}
where $\theta_{j}$ is the inclination and $c$ denotes a constant that is independent of inclination. 
Here the designations '1' and '2' refer to the two models.
Figure \ref{scaleplot} shows the deviation from constancy of the scale factors
 $c_{T_{\mathrm{eff}}}$ and $c_{L}$
for stellar models with masses between $1.875M_{\odot}$ and $3M_{\odot}$. The figure
was created by calculating the percentage difference between the constants at different inclinations 
and their values at $i=50^{\circ}$ for both plots.
These scaling relations become useful when trying to know
what the actual $T$ and actual $L$ are, as well as to provide surface properties 
such as and $T_{\mathrm{eff}}$, $g_{\mathrm{eff}}$, the surface radius $R$ and the rotational velocity $V$, as functions of latitude.

\articlefiguretwo{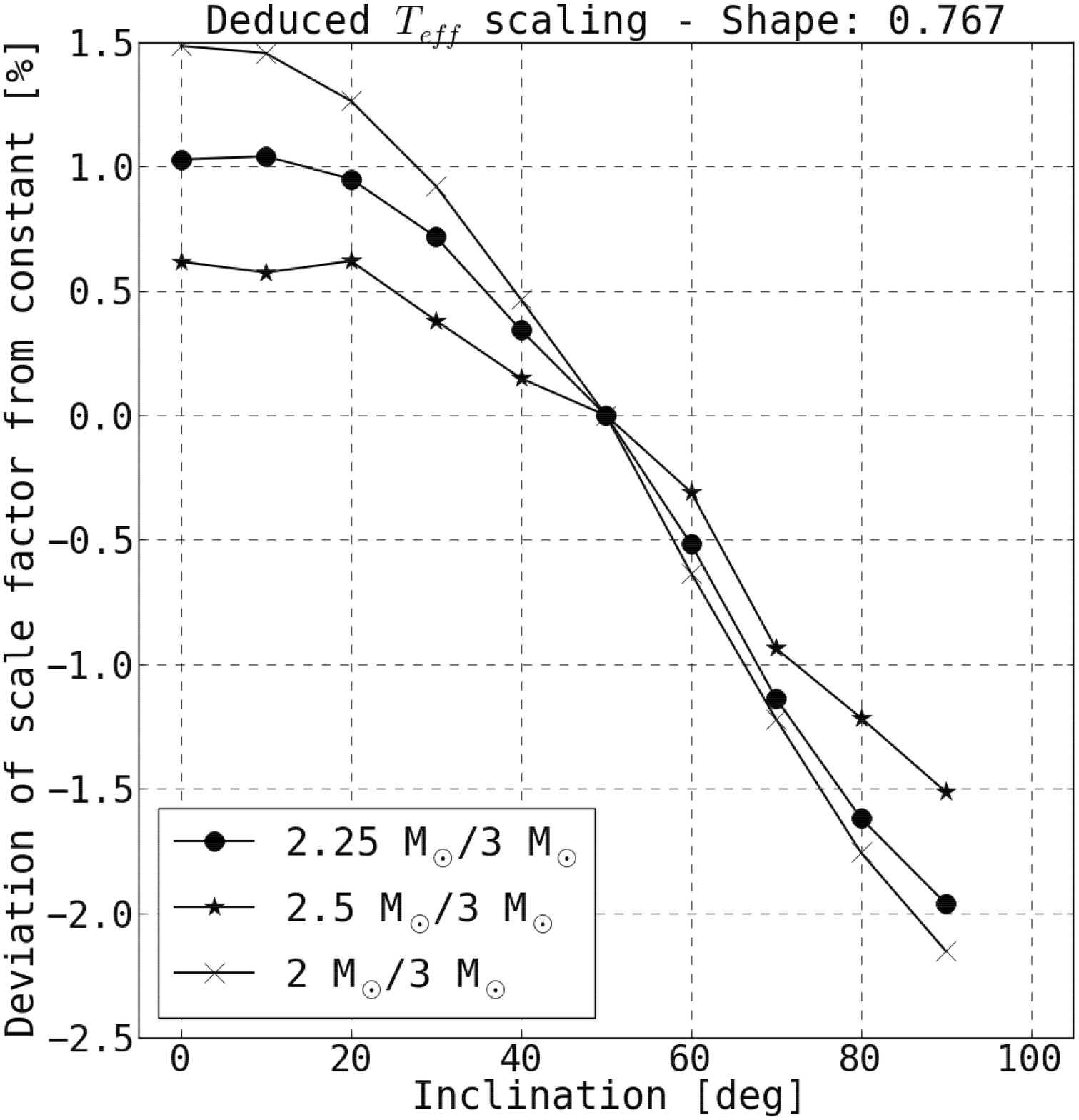}{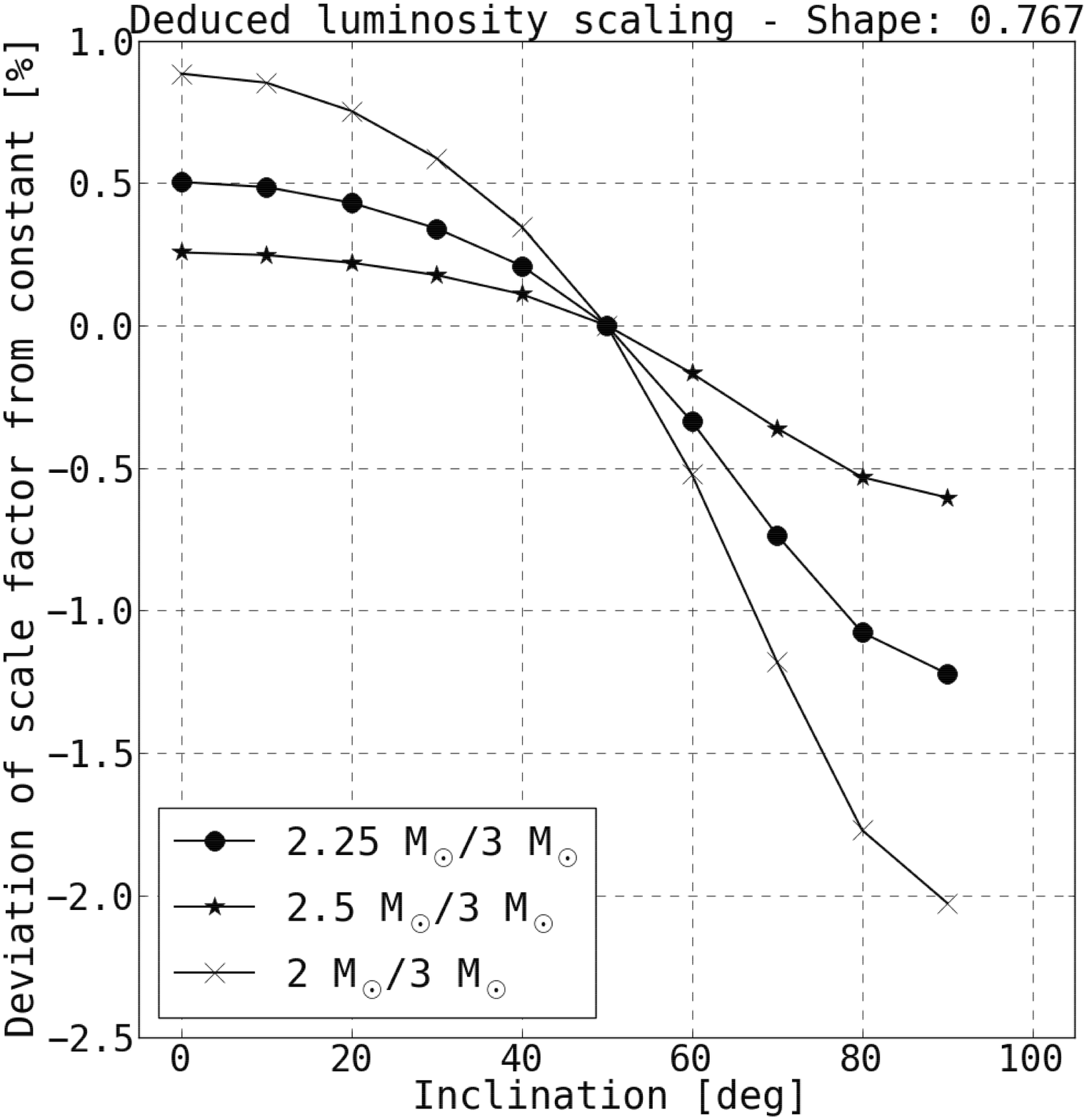}{scaleplot}{Constancy of scale factors $c_{T_{\mathrm{eff}}}$ and $c_{L}$  \emph{Left:} Deduced $T_{\mathrm{eff}}$  \emph{Right:} Deduced $L$}

There are some limitations, however, because some assumptions are required.
We have assumed that the surface is an equipotential, which only
exists if the rotation law is conservative. Even for conservative
rotation laws, it remains an assumption that the surface is an
equipotential. This likely matters for the part of the solution that
makes an estimate of the mass, but it need not affect the scaling
of the observable properties as long as whatever mechanism
determines the surface shape determines it in the same way for
both the unknown and comparison objects. Our very limited
knowledge of the surfaces of rotating stars does not allow an
answer to this question.

In order to find an application for these scaling relations, we found
in the literature rapidly rotating stars that have been
observed with interferometry. The next section will present the work done
on one of those rotating stars.

\section{Test Case: Vega}

Vega has been a photometric standard for a good part of the last century, but 
suggestions of anomalies in its luminosity, radius and shapes of the 
weak lines when compared to other A0 V type stars indicate that
Vega is not the perfect standard it was once thought to be.
First hints of Vega's
abnormal luminosity came from calibration studies of the H$\gamma$ equivalent width to 
to absolute magnitude relationship \citep{Petrie1964,Millward1985}. The radius was found
to be larger than expected \citep{HanburyBrown1967,Ciardi2001}. Finally, another important 
element of Vega's peculiarities that hinted the rotating nature of Vega was
the flat-bottomed shapes of its weak lines \citep{Gulliver1991}. Work done by \citet{Elste1992} and
\citet{Gulliver1994} showed that the shape of the lines could be explained if Vega was a
rapidly rotating star ($V_{\mathrm{eq}}=240\mathrm{km\,s}^{-1}$) viewed nearly pole-on ($i=5.5^{\circ}$).
Subsequent interferometric studies would confirm the rapidly rotating nature of Vega \citep{Peterson2006,Aufdenberg2006},
 and also establish that the star was rotating at nearly $V_{\mathrm{eq}}=270\mathrm{km\,s}^{-1}$ 
 ($\sim90\%$ of its breakup velocity).
Recent spectroscopic analysis, however, has lead to different results when trying to determine
the rotating properties of Vega. In particular \citet{Takeda2008} found that Vega had to be
rotating at a much lower rotational rate ($V_{\mathrm{eq}}\simeq175\mathrm{km\,s}^{-1}$) and at a
slightly higher inclination angle of $i\simeq7^{\circ}$ in order to match the observed spectral
features found on VEGA's SED.
Recently, findings of a periodic weak magnetic field in Vega \citep{Lignieres2009} allowed an independent
measurement of a rotating period of $P=0.71\pm0.03$ days \citep{Petit2010,Alina2012}, which is compatible with the period range
predicted by the line profile studies ($P\sim0.7-0.9$ days).
In light of these latest results \citet{Monnier2012} used the latest CHARA array data to obtain new interferometric
data for Vega, and after applying better calibration methods for the observations together with new rotating models
it was found that it is possible to get more slowly rotating models that match the interferometric observations.
It is important to note that part of the analysis in this work shows the strong dependence of the gravity darkening
parameter $\beta$ to the fit of the observed interferometric data. Depending on the value of $\beta$ it is possible
to find models that agree with both previous interferometric analysis as well as with the spectroscopic work that
suggest a slower rotating case.

We believe we can contribute to the analysis of Vega by having realistic 2D rotating models from which we can
apply the scaling relationships described before to find the appropriate surface properties of Vega that match the observed
interferometric data, SED and corresponding weak lines used in previous studies. A virtue of these models and this method is that
the relationship between $T_{\mathrm{eff}}$ and $g_{\mathrm{eff}}$ is based on a grey atmosphere
relation between the surface temperature and $T_{\mathrm{eff}}$, effectively removing the need to
arbitrarily impose a gravity darkening parameter $\beta$. Interestingly, calculated values of $\beta$
 for our models were compared with recent gravity darkening studies \citep{Espinosa2011}, finding similar results.

A first step to obtaining an appropriate model is the selection of target parameters of Vega. Table 1 shows the set of parameters
chosen to use the scaling relations. The method uses this information as input as well as the surface properties of pre-computed rotating
stellar models with the same $R_{\mathrm{p}}/R_{\mathrm{eq}}$, to return the surface properties 
($T_{\mathrm{eff}}$, $g_{\mathrm{eff}}$, the radius $R$ and the rotational velocity $V$) as
functions of latitude. A more detailed description of the procedure is given in \citet{Castaneda2014}.

\begin{table}[!ht]
\caption{Selected properties of Vega}
\smallskip
\begin{center}
{\small
\begin{tabular}{rlc}  
\tableline
\noalign{\smallskip}
Parameter & Value & Source\\
\noalign{\smallskip}
\tableline
\noalign{\smallskip}
Deduced $T_{\mathrm{eff}}$ & $9550$K & SED \\
$i$ & $5^{\circ}$ & Interferometry \\
$V\,\mathrm{sin}\,i$ & $21.9\mathrm{km\,s}^{-1}$ & Line profiles \\
$R_{\mathrm{eq}}$ & $2.728R_{\odot}$ & Interferomtry and parallax \\ 
$R_{\mathrm{p}}$ & $2.20R_{\odot}$ & Assumed\\ 
\noalign{\smallskip}
\tableline\
\end{tabular}
}
\end{center}
\end{table}

With the surface parameters obtained, it is possible to calculate a synthetic image of the star
in the sky, which at the same time allows the calculation of synthetic interferometric
data to compare with observed information. The procedure is the same as the one described
by \citet{Aufdenberg2006}. We used this to compare how well the scaled model fit the data
used by \citet{Monnier2012} and the result is shown in Figure \ref{visplot}. The result is
comparable to the recent studies, but it goes against the recent findings that suggest that Vega is rotating at
a much slower rate.

\articlefigure{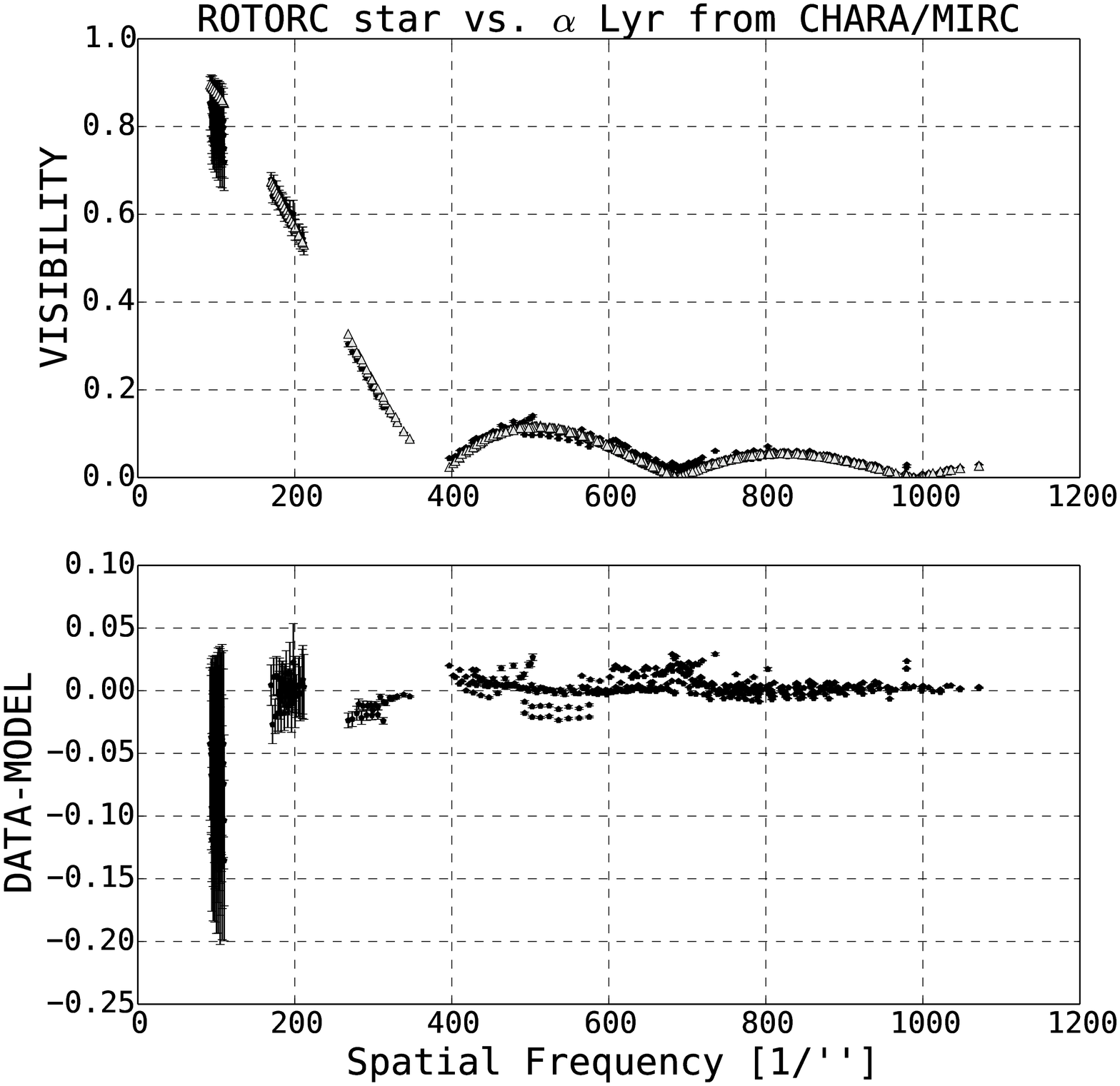}{visplot}{Comparison between the observed visibility curve of Vega and a synthetic visibility curve obtained by scaling the properties of a reference rotating model}

After the scaled surface parameters are obtained, one can compute an SED by performing
a weighted integral over all the contributions of the intensities
from every point on the stellar surface visible to the observer to obtain
the flux the observer would see. This approach for calculating the SED has been frequently used
(e.g., \citealt{Slettebak1980,Linnell1994,Fremat2005,Aufdenberg2006,Gillich2008,Yoon2008,Dall2011});
the specific details of our calculations are outlined by \citet{Lovekin2006a}.
We then compared some of the lines considered in
the analysis made by \citet{Takeda2007}. Figure \ref{linesplot} shows comparison of four selected weak lines.
Like other authors \citep{Takeda2008,Yoon2008,Yoon2010} have done in previous studies, we renormalized the line depth
by multiplying by an arbitrary factor to make the theoretical line strength consistent with the observed one.
This can be done because the weak lines involved (their maximum depth $\sim1\%$ ) are on the linear part of the curve of growth so their line depth should scale closely with $\epsilon gf$ 
(where $\epsilon$ is the abundance, g is the statistical weight, and f the is oscillator strength). These quantities depend on the atomic physics modelling and do not carry information about
the rotating properties of the star, which is the information we want to compare.

\articlefigurefour{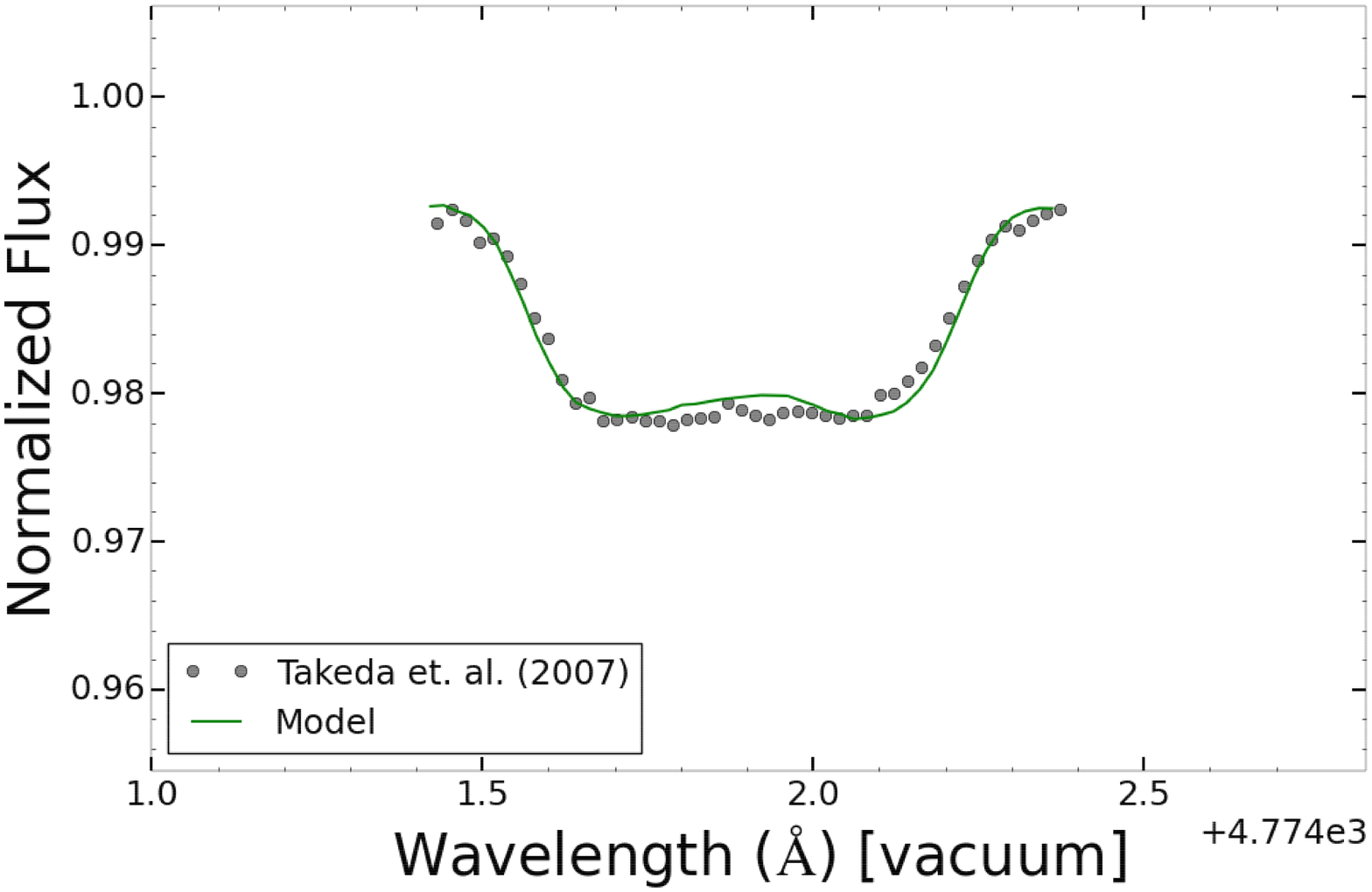}{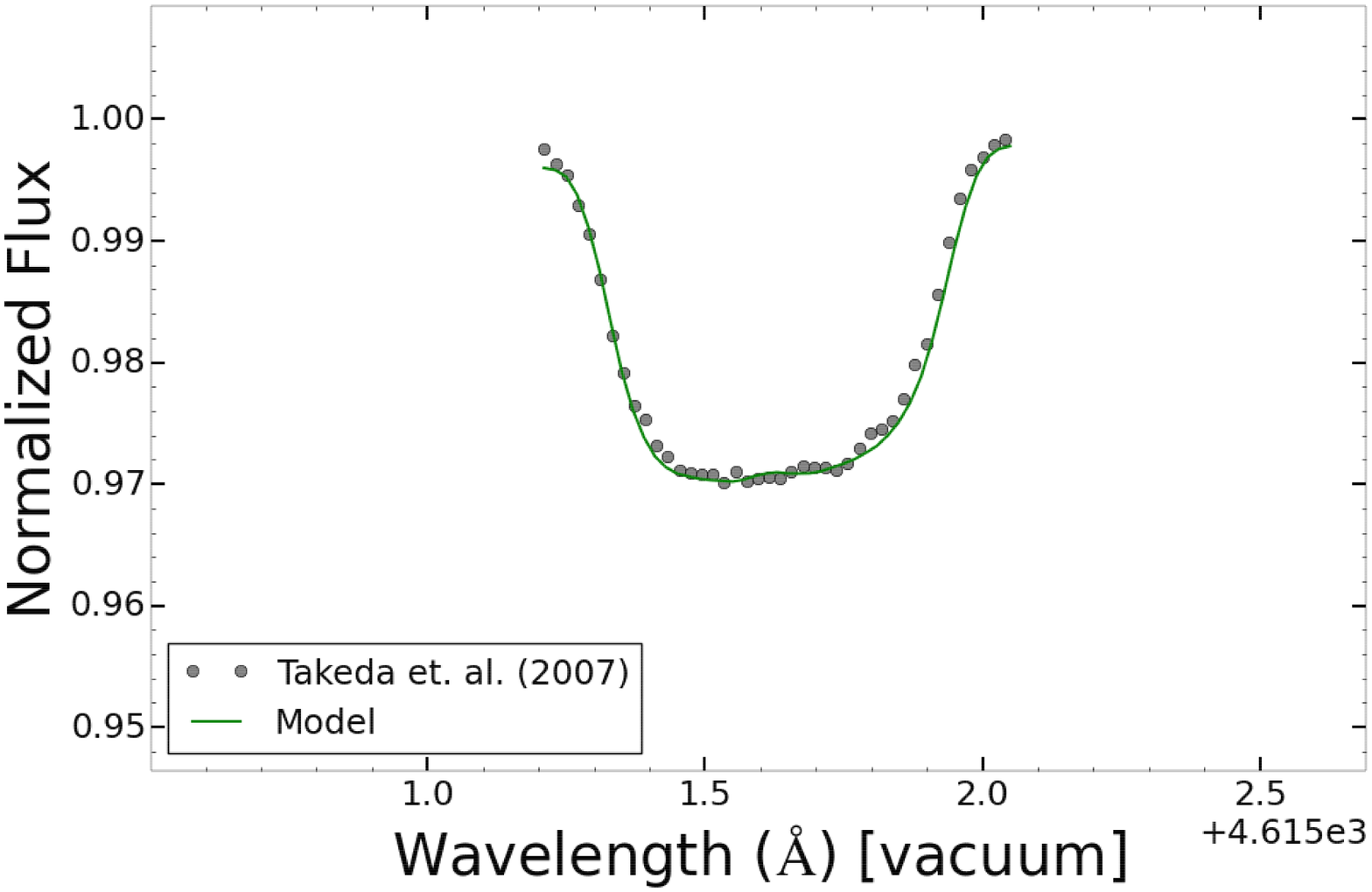}{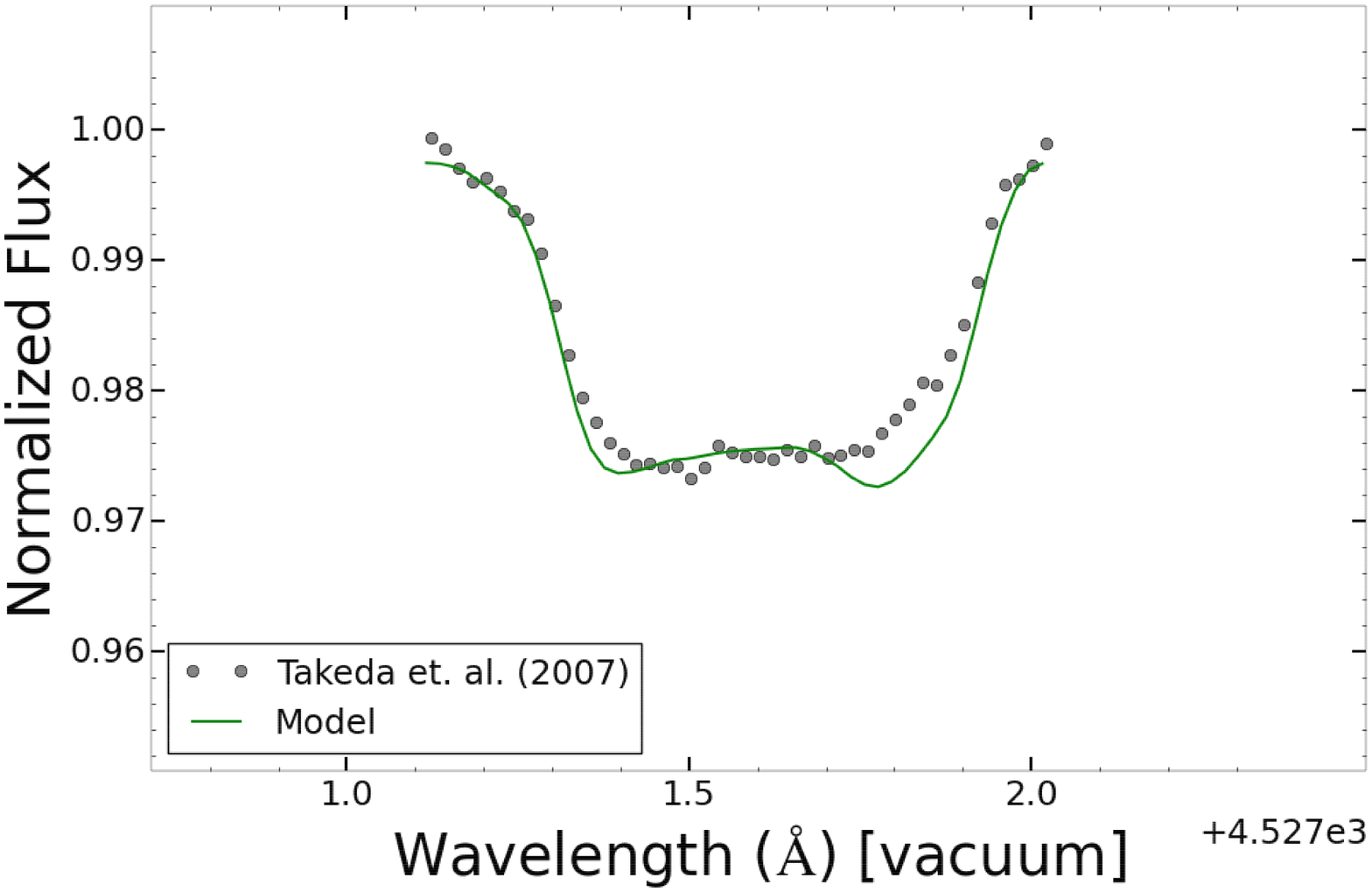}{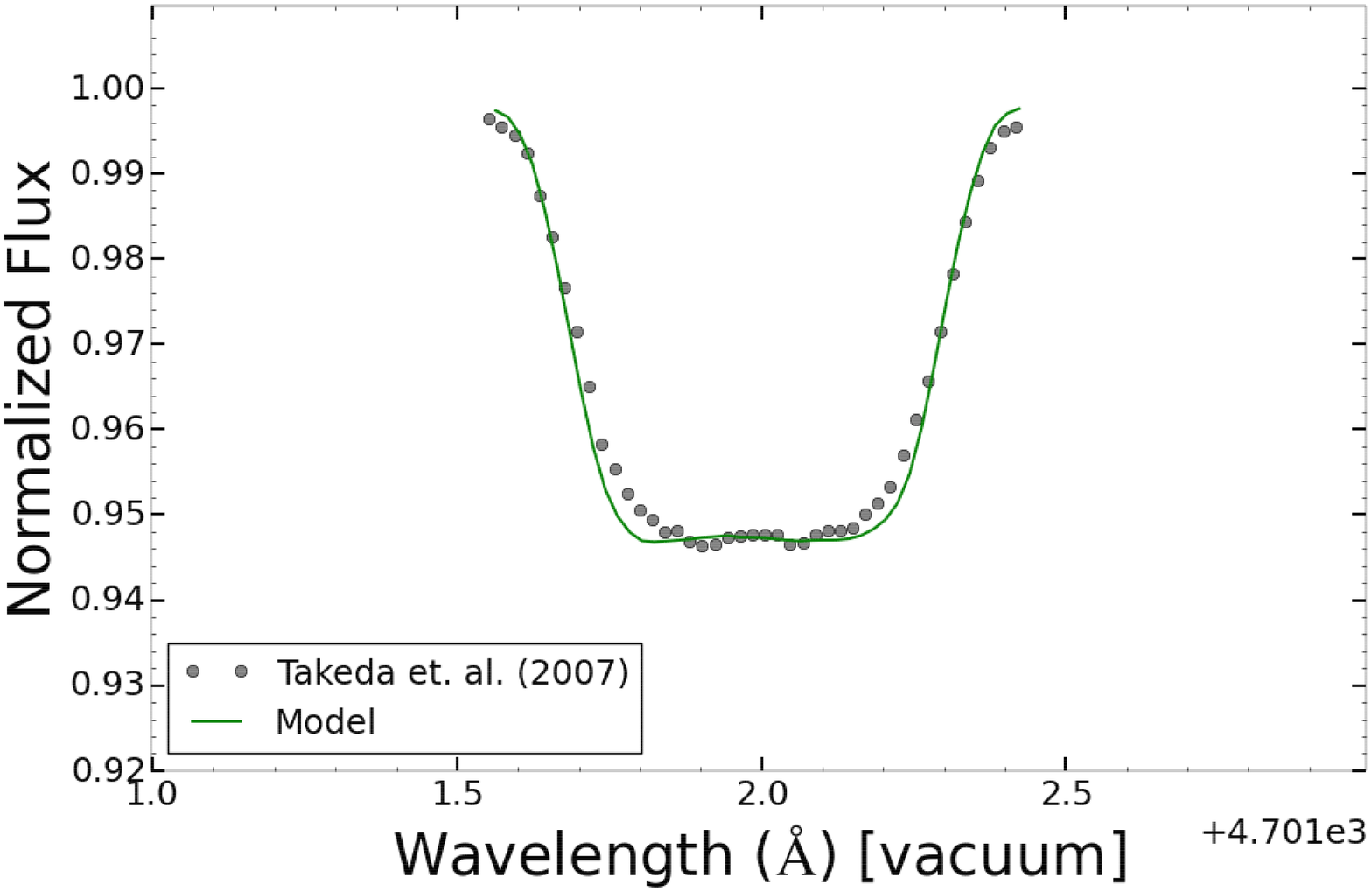}{linesplot}{Selected weak line comparison: \emph{Top Left}: CI-4775.8\AA, \emph{Top Right}: CrI-4616.6\AA, \emph{Bottom Left}: FeI-4528.6\AA, \emph{Bottom Right}: MgI-4703\AA}

\citet{Takeda2008} considered 196 weak lines. We compared our results with about 20 of them, finding similar agreement in the shape of the line profiles.
Although there are small discrepancies in the shapes of some lines, this is an interesting result
 because the fit is achieved using the model that matches the higher rotational velocity
parameters.

So far we have only performed this test by using one set of accepted observed properties of Vega.
A next step would be to consider a broader range of parameter space and document how different
results we get in those cases.

\section{Conclusion}

Interferometric observations combined with traditional observations provide a great opportunity to test
realistic rotating stellar models. With this in mind, the preliminary analysis of Vega using the scaling 
relationships of observable properties in rotating stars is used to translate what we
observe from a rotating star to physically useful information under the appropriate conditions.
The study of the Be phenomenon, although more complex because of the presence of the disk, could benefit from this technique if
one can know both the inclination and the surface shape of the star.

\acknowledgements We would like to thank John Monnier for providing us with the visibility data for Vega. We also thank Compute Canada and ACEnet for the computational resources used in this research.

\bibliography{references}  

\subsection*{Questions asked after the talk}

\textbf{Q:} \textit{"Maybe you’re aware that Vega is the prototype
of a possibly large class of stars in which very weak surface magnetic
fields are present. How do your derived period and radius compare with
those inferred from Zeeman Doppler Imaging by Petit et al.?"}

\noindent \textbf{A:} Please refer to Page 4, line 14 in this document.

\noindent \textbf{Q:} \textit{"You mentioned that your models allow for latitudinal variation of
gravity darkening that can differ from von Zeipel’s law. Could
explain more what more general conditions are used?"}

\noindent \textbf{A:} A comment about this aspect of our models
was made in page 4: End of first paragraph
 and second paragraph.

\end{document}